

 \magnification=1200
 \vsize=23.5 truecm
 \hsize=16 truecm

\nopagenumbers\parindent0pt\parskip13pt
\headline={\ifodd\pageno\rhl \else \lhl\fi}
\def\rhl{\tenrm\hfill\ -- \folio --}
\def\lhl{\tenrm-- \folio --\hfill}
\overfullrule 0pt

\def\lsim{\raise0.3ex\hbox{$<$\kern-0.75em\raise-1.1ex\hbox{$\sim$}}}
\def\gsim{\raise0.3ex\hbox{$>$\kern-0.75em\raise-1.1ex\hbox{$\sim$}}}


\def\half{\hbox{ ${1\over 2}$ }}
\def\third{\hbox{ ${1\over 3}$ }}
\def\fr#1#2{{\scriptstyle{#1\over#2}}}
\baselineskip 16pt
\pageno=0
\line{\hfill Preprint HU-TFT-92-1}
\line{\hfill 8 January 1992}
\line{\hfill revised 9 March 1992}
\line{\hfill re-revised 22 April 1992}
\vskip 2.5cm
\centerline{THE FREE ENERGY OF SPHERICAL BUBBLES}
\centerline{IN LATTICE SU(3) GAUGE THEORY}
\vskip 1cm
\centerline{K. Kajantie$^1$, Leo K\"arkk\"ainen$^{2}$
and K. Rummukainen$^3$}
\vskip 1.5cm
\centerline{Abstract}
\vskip 0.6cm
We study the coefficients of the expansion $F(R)=\third c_3R^3 +
\half c_2R^2+ c_1R$ of the free energy of spherical bubbles at $T=T_c$
in pure glue QCD using lattice Monte Carlo techniques. The coefficient
$c_3$ vanishes at $T=T_c$ and our results suggest that the sign and
the order of magnitude of $c_1$ is in agreement with the value
$c_1=\pm 32\pi T_c^2/9$ (- for hadronic bubbles in quark phase, + for
quark bubbles in hadronic phase) computed by Mardor and Svetitsky from
the MIT bag model.  The parameter $c_2$ is small in agreement with
earlier determinations.

\vfill
\hrule\medskip
1) University of Helsinki, Department of Theoretical Physics,
Siltavuorenpenger 20 C, 00170 Helsinki, Finland;
kajantie@finuhcb.

2) Research Institute for Theoretical Physics,
Siltavuorenpenger 20 C, 00170 Helsinki, Finland; permanent address
Universit\"at Bielefeld, Fakult\"at
f\"ur Physik, Postfach 8640, D-4800 Bielefeld 1, BRD;
leo@c240.uni-bielefeld.de.

3) CERN/TH, CH-1211 Geneve, Switzerland; rummukai@cernvm.

\eject
\pageno =1
\null\vskip 0.6cm
Bulk QCD matter is stable in the quark-gluon plasma phase (Q)
for $T\ge T_c$ and in the hadronic phase (H) for $T\le T_c$. At
$T=T_c$ the free energies coincide and there is a latent heat $L$:
$$
F_q(T_c)= -Vp_q(T_c) =F_h(T_c)= -Vp_h(T_c),\quad L=T_c
[p^\prime_q(T_c) -p^\prime_h(T_c)], \eqno(1)
$$
assuming the transition is of first order and taking the chemical
potential $\mu=0$. The determination of the free energy has been studied
in great detail with lattice Monte Carlo techniques [1]. At $T=T_c$ a
stable H-Q (order-disorder) interface can exist in the system and
contributes to the free energy:
$$
F=-pV+ \alpha A. \eqno(2)
$$
Lattice Monte Carlo computations [2-3] using planar interfaces have
given the value
$$
\alpha(T_c) \approx 0.1T_c^3 \eqno(3)
$$
for the interface tension in the case of pure glue.

The purpose of this letter is to go still one step further in the
lattice Monte Carlo determination of free energies of domains of one
phase immersed in the other: spherical bubbles of H phase in Q matter
(or Q bubbles in H matter) [4]. Then the free energy of an H bubble
(relative to a system entirely in Q phase) with radius $R$ can be
expanded in powers of $1/R$ as follows:
$$
F(T,R)=[p_q(T)-p_h(T)]\fr43\pi R^3 +\alpha(T)4\pi R^2+\gamma(T) R+\dots
\eqno(4)
$$
The new element here is the last curvature term. Using the MIT bag
model, i.e., treating the bubble as a spherical cavity, gluons as
eight copies of Abelian photons, and assuming bag boundary
conditions, Mardor and Svetitsky [4] found that
$$
\alpha_{\rm bag}(T)=0,\quad \gamma_{\rm bag}(T)=-{32\pi \over9}T^2;
\eqno(5)
$$
for Q bubbles in H the sign of $\gamma$ is positive. Our
main result is that also lattice Monte Carlo shows
evidence of the pattern in eq. (5): a small area term $\alpha$
and a large curvature term $\gamma$, the numerical value of which
is compatible with that in eq.~(5).

If only the terms shown in (4) contributed to $F(R)$, there would be
an $R_{\rm min}$ with $F(R_{\rm min})<F(0)=0$, i.e., H bubbles would
be stable even for $T> T_c$. However, even in the bag model
the terms in $1/R$ omitted in eq.~(4) decrease $F(0)-F(R_{\rm min})$
and on the basis of our results presented here one
cannot conclude anything on the stability of H bubbles for $T>T_c$.

\bigskip
Before presenting the lattice results, consider eq.~(4)
phenomenologically for $T\approx T_c$. Scaling with $T_c$: $\hat
T=T/T_c, \hat R=T_cR$,  we firstly can write $p_q(T)-p_h(T)= L(\hat T
-1)$ and estimate the latent heat by $L=T_cp^\prime_q(T_c)=
32\pi^2T_c^4/45$. For H-Q interfaces we do not know $\alpha(T)$ away
from $T_c$, but a study of order-order interfaces in the Q phase [5]
has given a very large value, $T\alpha'(T)/\alpha(T)\approx30$, for
the surface entropy, $S_s=-\alpha'(T)A$. We use this same value and
write $\alpha(T)/T_c^3 = 0.1+3(\hat T-1)$. Our lattice data below
indicates that $\gamma$ depends only weakly on $T$ and we
finally have
$$
F(T,R)/T_c= 4\pi \biggl\{ {32\pi^2\over135}(\hat T-1)\hat R^3 +
[0.1+3(\hat T-1)]\hat R^2 -{8\over9}\hat R \biggr\}. \eqno(6)
$$
Quantitatively, this model implies that at $T=T_c\approx200$ MeV a
stable H bubble has a radius of about 4/$T_c$=4 fm. The radius has
shrunk to 1 fm at $T=1.05T_c$.

Our aim is to determine the coefficients of the expansion (4) at
$T=T_c$.  First, we point out that the expected magnitude of
$\hat\gamma = \gamma/T_c^2 = -32\pi/9$ is much larger than
$\hat\alpha=\alpha/T_c^3 = 0.1$.  This difference is
essential for the success of our calculations: as it will turn out,
the actual thickness of the bubble wall is of order $1/T_c$, so that
the {\it definition} of the radius of the bubble is necessarily
ambiguous.  For example, possible non-equivalent choices are the
radius where the order parameter reaches some fixed value between the
ordered and disordered states; the turning point of the radial
order parameter; or the corresponding quantities for the energy
density.

This ambiguity affects the functional form of $F(R)$. Let us redefine
the radius $\hat R\to\hat R_1 =\hat R-\delta$, so that at $T=T_c$:
$$
\hat F = 4\pi\hat\alpha\hat R^2 + \hat\gamma \hat R
       = 4\pi\hat\alpha\hat R_1^2 + (\hat\gamma +
8\pi\hat\alpha\delta)\hat R_1 + {\cal O}(R_1^0) .\eqno(7)
$$

If now $\hat\alpha$ and $\hat\gamma$ are of the same magnitude, even a
slight variation in $\delta$ will spoil the ${\cal O}(R_1)$ term.  In
fact, in this case a prime candidate for the bubble radius would be a
value of R where $F = 4\pi\alpha R^2$ exactly!  However, the values
given above for $\hat\gamma$ and $\hat\alpha$ imply that $\delta$
would have to be $\approx 4/T$ for the linear term to vanish.  This is
more than the wall thickness $W\sim 2/T$, so that this $\delta$ would
move the radius completely outside of the actual interface.  Thus, the
whole concept of the curvature contribution to the free energy makes
sense only when $\gamma\gg 8\pi\alpha W \approx 8\pi\alpha/T$.

We would like to stress that this ambiguity in the subleading terms is
of quite general nature; it is present both in physical phenomena and
in lattice calculations.  We shall return to this question more in
detail further down.

On a homogeneous system at thermal equilibrium the bubbles are rare and
lattice calculations would be extremely time-consuming.  To remedy
this and to gain systematic control of the bubbles, we choose a
spatially spherical region of radius $R$ and adjust the couplings
$\beta\equiv 2N_c/g^2 =\beta_c\pm\Delta\beta$ in the interior and
exterior, respectively.  This acts like an external field and
stabilises the bubble.  For simplicity, we call this $R$ the radius of
the bubble, and check the sensitivity of the results to
reparametrisations later.  The coefficients of eq.~(4) are now
evaluated with various $\Delta\beta$, and the final answer is obtained
by extrapolating $\Delta\beta\to 0$.  In the calculations one cannot
use too small a $\Delta\beta$, otherwise the bubble will vanish.  We
assume that the relevant quantities behave linearly in $\Delta\beta$,
as long as the bubble is preserved.

It is not possibly to evaluate directly the free energy $F$ using a MC
simulation. However, the derivatives of the free energy are accessible
as ensemble averages, thus providing, after integration, the value of
the free energy up to a constant.  The `standard' way to do this is to
integrate with respect to the inverse coupling $\beta$ [1,2,3].
However, as we are interested in the $R$-dependence of the free
energy, it is natural to take the derivativee w.r.t.~$R$, the bubble
radius.  With this method we avoid the very delicate and
time-consuming integration over $\beta$. However, in order to measure the
derivative, we have to modify the action to form
$$
\sum_{\rm n} \beta (r_{\rm n},R) S(n) =
\sum_{\rm n}
(\beta_c + \Delta\beta\tanh {r_{\rm n}-R\over w} )
S_{\rm Wilson}(n)  , \eqno(8) 
$$
where the sum is over the lattice, $r_{\rm n}$ is the distance
from the center of the bubble and $w$ (chosen as 0.5 lattice spacings
$a$) parametrises the width of the cross-over of the function
$\beta(r,R)$; $w$ is not to be mixed with the actual thickness of the
interface. The quantity accessible by means of MC then is
$$
{ d(F/T)\over dR } = - { \Delta\beta \over w }
\left \langle \sum_{\rm n} \cosh^{-2}( {r_{\rm n}-R\over w})
S(n) \right \rangle, \eqno(9)
$$
where the average is taken over the ensemble created with
the action of eq.~(8).

The role of the parameter $w$ deserves some elaboration. It was
introduced to facilitate the evaluation of the derivative
$d\beta(r_{\rm n},R)/dR$: the most obvious choice for $\beta(r,R)$
would be a step-like function ($w=0$), but then eq.~(9) becomes
$-2\Delta\beta \sum_{\rm n} \delta(r_{\rm n} - R)S(n)$.  On a 3+1
dimensional lattice this clearly puts different values of $R$ to a
completely unequal footing, and the whole operator makes sense only
when one integrates it over some range of $R$.  In principle this
requires separate simulations for each value of $R=r_{\rm n}$ in this
range, which would be exceedingly costly.  The form of $\beta(r,R)$ in
eq.~(8) smears the step over a shell $R-w \lsim r_{\rm n} \lsim R+w$.
The larger $w$ is, the smoother one can expect the behavior of $F$ as
a function of $R$ to be, and $F'(R)$ measured at any single value of
$R$ becomes meaningful.  However, too large a $w$ does not pin down the
physical surface well enough.

We chose $w=0.5a$ in order to keep the cross-over in $\beta(r,R)$
within one lattice spacing, which is anyway the shortest length scale
on the lattice.  It turns out that this choice makes the function
$F'(R)$ sufficiently regular (although there still remains some
systematic lattice effects, as seen from the similarity of the
fluctuations of the data around the fits in Fig.~2.).

We stress that the use of the $w=0$ instead of non-zero $w$ does not
make the calculations any more physical: when $\Delta\beta\to 0$, the
explicit $w$-dependence of the action vanishes, and the configurations
extrapolate to physical bubble configurations.  One can also expect
that the effects of a particular choice of $w$ become minimal.  As
discussed above, the bubble radius can be defined in numerous ways,
and the discontinuity (or the cross-over) radius of the `external
field' $\Delta\beta(r,R)$ is not particularly good parametrisation,
since it is not any intrinsic property of the bubble.  For example, it
will turn out that the radius where the order parameter reaches the
value half-way between the values in the interior and exterior bulk
states can deviate from the cross-over radius considerably more than
the half-width of the cross-over itself (Fig.~1).  This implies that
even with finite $\Delta\beta$ the effects caused by choosing $w=0$ or
$w=0.5$ are overshadowed by the difference between the `intrinsic'
radius and the `external field' radius of the bubble.  We remind again
that this problem is only alleviated by the large value of the
curvature term when compared to the interface tension.

On the lattice we actually measure $R$ in
units of the lattice spacing $a=1/(N_tT)$, and have, from eq.~(4),
$$
{d(F/T)\over R/a}= {4\pi\over N_t^3}{p_q(T)-p_h(T)\over T^4}
{R^2\over a^2} + {8\pi\over N_t^2}{\alpha(T)\over T^3} {R\over a}
+{1\over N_t}{\gamma(T)\over T^2} +\dots. \eqno(10)
$$
Denoting the LHS by $F'(R)$ we fit the lattice data to (from now on
$R/a\to R$)
$$
F'(R) = c_3 R^2 + c_2 R + c_1, \eqno(11)
$$
extrapolate the coefficients $c_i$ to $\Delta\beta=0$ and obtain the
physical quantities at $T_c$ by comparison with eq.~(10).


We have performed simulations on a 2$\times16^3$ lattice using an
algorithm for SU(3) gauge theory with a combination of
pseudo-heat bath and overrelaxation sweeps. The values of
$\beta$ were $\beta_{\rm in/out}= 5.1\pm\Delta\beta$ with
$\Delta\beta=0.5$, 0.25, 0.125 and the radius R varied between 2 and 7
in steps of 0.5 with some intermediate points in steps of 0.25. The
total number of iterations thus was 2(for H,Q)$\times$
47(for values of $\Delta\beta$ and $R$)$\times$10000 (or sometimes
5000) iterations per point, amounting to a total of about 1M iterations
or (with 2.7 s/sweep) about 700h of Cray XMP CPU time.
By modern standard, the number of iterations on each point is relatively
low, but the systems are rather far from the critical coupling and the
autocorrelation times are small, $\lsim 10$ sweeps.

The size of the lattice limits the physical bubble radius to $\le
3.5/T$.  This is a serious limitation; it forces us to use relatively
large $\Delta\beta$ to maintain the bubble on the lattice and to keep
the correlation length small enough.  Using eq.~(6), one can see that
at $T=T_c$ the expected stable radius is $4/T_c$ and the radius where
$F(R) = F(0) = 0$ is $9/T_c$.  To truly study this region, the spatial
size of the lattice should be at least $\approx 40a$.  The small size
of the bubbles and large $\Delta\beta$ values can cause systematical
errors, whose magnitude remains unknown; for example, the terms ${\cal
O}(R^n)$, $n\le 0$ in eq.~(4) can become significant.  Also, for a
more detailed analysis of the $\Delta\beta$ dependence one would need
more than 3 values of $\Delta\beta$ for both Q and H bubbles.

Fig.~1 shows how the order parameter varies as a function of the
distance from the center of the bubble for $\Delta\beta=0.25$; the
pattern is similar for $\Delta\beta=0.5$ or 0.125. One sees clearly
the formation of an H bubble but with an interface thickness of some 4
units in $R$. In physical units the interface would be about 2 fm
thick: we have $1/T_c=N_t a\approx$ 1 fm, so that $a\approx0.5$ fm.
Remember that the input thickness $w$ was only half a lattice spacing.

The results for $F'(R)$ and for the fitted parameters $c_i$ are shown
in Figs.~2 and 3 for H bubbles in Q and in Figs. 4 and 5 for Q bubbles
in H.

Since the numerical values of $c_3R^2$ are rather large, we shown in
Fig.~2 the measured values of $F'(R)$ after subtracting the
term $c_3R^2$ with the fitted values of $c_3$ shown in Fig.~3(a). Only
values $R\ge3$ are used in the fit. The linear
behaviour of the data is evident, although the $\chi^2$/d.o.f. is
as large as 2-4 depending on the lower cutoff in $R$. We attribute the
large value of $\chi^2$/d.o.f. to a lattice discretisation effect: the
number of plaquettes inside the bubble does not grow smoothly with
increasing $R$.

The values of $c_3$ in Fig.~3(a) extrapolate linearly to
-0.38$\pm$0.10 when $\Delta\beta\to0$. Relative to the values
extrapolated from, about -80,-40,-20, this is very close to the value
0 expected at $T=T_c$ (eq.~(4)).  Also the values for
$\Delta\beta\not=0$ are compatible with the pressure difference
between homogeneous simulations at $\beta_{\rm out}$ and $\beta_{\rm
in}$. Similarly, $c_2=2\pi\alpha/T^3$ (eq.~(10)) extrapolates to the
value 0.63$\pm$0.89 when $\Delta\beta\to0$.  This coincides with the
value (3) of $\alpha(T_c)$, but due to the large statistical errors we
can confirm only the smallness of this value.  Obviously, the methods
utilised in [2,3] for planar interfaces are superior to the one used
here for determining the interface tension $\alpha$.  The values of
$c_2$ for $\Delta\beta\not=0$ are discussed below.

The main result is shown in Fig.~3(c): the curvature parameter
$c_1=\gamma/(N_tT^2)$ as given by the intercept of the lines in
Fig.~3(a) at $R=0$. It is independent of $\Delta\beta$ and well
compatible with the bag model result.  The errors are of the same
magnitude as with $c_2$; however, due to the large value of $c_1$
the result is $3\sigma$ away from zero.

The corresponding data and fits for Q bubbles in H presented in
Figs.~4 and 5 show a similar average pattern, but the error bars in
the fits are so large that they make the result only $1\sigma$ away
from zero.  In spite of this, the central value of $c_1$ coincides
with that obtained for H bubbles, but has an opposite sign.  Due to
the large errors in Q bubbles, in what follows we concentrate mainly
on the H bubble case.

{}From the derivative $F'(R)$ we can obtain $F(R)$ by integration. The
result is shown in Fig.~6, normalised arbitrarily to $F(0)=0$. One sees
clearly the large numerical values for $\Delta\beta\not=0$ caused by the
dominant $R^3$ term. The distance of extrapolation to $\Delta\beta=0$ is
evidently large and we have constrained $c_3(\Delta\beta=0)=0$. The
resulting curve is in agreement with the phenomenological one in
eq.~(6). Notice, however, that one should not take the $1/R$ expansion
too seriously for very small $R$.

As discussed above, the definition of the physical radius of the
bubble is necessarily ambiguous.  It is essential to check the
stability of the results against these reparametrisations. Let us
perform a reparametrisation of $R$ consistent with the $R\to\infty$
behaviour:
$$
R\to R+x_1+x_2/R+{\cal O}(1/R^2). \eqno(12)
$$
Then
$$
F(R)= \third c_3R^3 + \half(c_2+2c_3x_1)R^2 +
[c_1+c_2x_1+c_3(x_1^2+x_2)]R +
{\cal O}(R^0). \eqno(13)
$$
We are mainly interested in the term linear in $R$. In it $c_3$ is
large but vanishes for $\Delta\beta\to 0$ and thus does not affect the
physical value of $c_1$. In the remaining term we know $c_2$ from the
value of the interface tension but $x_1$, which corresponds to a
constant shift in the radius, remains a free parameter.  Even if we
adjust it by $\sim \pm 1$ lattice units, which should encompass all
`reasonable' radius redefinitions discussed above, the linear term in
$R$ is only modified by $\sim 10\%$.  This is well below the
statistical errors of $c_1$.

It is also possible to fix $x_1$ by comparing the second term with the
values of the area term in planar interface simulations, i.e., the
data in Fig.~3(b) with the data in Fig.~8 of [2]. One obtains
$x_1=0.04$ independent of $\Delta\beta$, which shows the consistency
of the discussion. Thus at $\Delta\beta=0$ the value of $c_1\approx
-6$ is only modified by a term of the magnitude 0.03; totally
negligible.

As discussed above, this seemingly surprising stability of $c_1$ is
caused by the vanishing coefficient $c_3$ when $\Delta\beta \to 0$ and
by the very small value of the interface tension $c_2$ when compared
with $c_1$.  For $\Delta\beta\not=0$ the additional term $2c_3x_2$ may
explain why the data for $c_1$ seem independent of $\Delta\beta$ --
clearly it is also possible that $\gamma(T)/T^2$ is only weakly
dependent on $T$, as the bag model predicts.

In conclusion, our lattice Monte Carlo simulations support the
existence of the curvature contribution $\gamma$ to the free energy of
spherical bubbles predicted by the MIT bag model [4].  The value
predicted by the bag model is $\gamma/T^2=\pm 32\pi/9$, which is the
same order of magnitude than our results; although for the quark
bubbles the result is not statistically significantly non-zero.
However, one should emphasise that the lattice used has a small
$N_t=2$, that even the smallest value of $\Delta\beta$ still is quite
large, that the bubble radii accessible to us may be too small to
justify the truncation of the $1/R$ -expansion used here, and
that there is no control of finite $V$ effects.  It is quite
conceivable that more refined analyses will lead to a different
numerical value, but we nevertheless find it quite interesting that
Monte Carlo computations give such a clear signal for the MIT bag
model value for the curvature term. For the interface tension $\alpha$
we could not obtain a statistically significant result.  With further
Monte Carlo runs one should also be able to make statements concerning
$F(R)$ when $R\to0$ and to address the question of stability of H
bubbles for $T>T_c$ quantitatively.

\vskip1cm
\line{\it Acknowledgements\hfill}
We thank the Finnish Center for Scientific Computing for computing
facilities, the Academy of Finland for financial support and Jean
Potvin, Ben Svetitsky and Frithjof Karsch for discussions and
comments.
\vskip1cm

\vfill\eject
\null\vskip 1cm
\centerline{\bf References}
\vskip 0.6truecm\parindent15pt
\item{[1]} J. Engels, J. Fingberg, F. Karsch, D. Miller and M. Weber,
Phys. Lett. B252(1990) 625.
\item{[2]} K. Kajantie, Leo K\"arkk\"ainen and K.
Rummukainen, Nucl. Phys. B333 (1990) 100.
\item{[3]} S. Huang, J. Potvin, C. Rebbi and S. Sanielevici,
Phys. Rev. D42 (1990) 2864.
\item{[4]} Israel Mardor and Ben Svetitsky, Phys. Rev. D44(1991)878.
\item{[5]} K. Kajantie, Leo K\"arkk\"ainen and K.
Rummukainen, Nucl. Phys. B357(1991)693.
\parindent0pt
\vfill\eject
\null\vskip 1cm
\centerline{\bf Figure Captions}
\bigskip
Fig.~1. The behaviour of the order parameter, the Wilson line $L$, as
a function of the distance $r$ from the center of the H bubble for
$\Delta\beta=0.25$ and for $R=3,4,5,6,7$. The dotted line shows
$\beta(r,R=5)$.
\medskip
Fig.~2. $F'(R)-c_3R^2$ for H bubbles. From up to down,
$\Delta\beta=0.5,0.25,0.125$; the straight lines are the fits
$c_2R+c_1$.
\medskip
Fig.~3. The fitted values of (a) $c_3$, (b) $c_2$, (c) $c_1$ as a
function of $\Delta\beta$ for H bubbles with a linear fit to the data
points and an extrapolation to $\Delta\beta=0$. In (c) the arrow shows
the bag model point -16$\pi$/9.
\medskip
Fig.~4. $F'(R)-c_3R^2$ for Q bubbles. From up to down,
$\Delta\beta=0.5,0.25,0.125$; the straight lines are the fits
$c_2R+c_1$.
\medskip
Fig.~5. The fitted values of (a) $c_3$, (b) $c_2$, (c) $c_1$ as a
function of $\Delta\beta$ for Q bubbles with a linear fit to the data
points and an extrapolation to $\Delta\beta=0$. In (c) the arrow shows
the bag model point 16$\pi$/9.
\medskip
Fig.~6. The free energy $F(R)/T$ at $T=T_c$, normalised to $F(0)=0$,
integrated from the measured values of $F'(R)$. $R$ is in units of $a$
and the curves are from below for $\Delta\beta=0.5,0.25,0.125$ and for
the extrapolation to $\Delta\beta=0$.

\bye